\documentclass[twocolumn,showpacs,aps,prl,superscriptaddress]{revtex4}

\usepackage{graphicx}
\usepackage{dcolumn}
\usepackage{amsmath}
\usepackage{epsfig}

\input babarsym

\newcommand{\BABARPubYear}    {05}
\newcommand{\BABARPubNumber}  {005}

\newcommand{\SLACPubNumber} {11054}

\def\figurebox#1#2#3{%
    \def\arg{#3}%
    \ifx\arg\empty
    {\hfill\vbox{\hsize#2\hrule\hbox to #2{\vrule\hfill\vbox to #1{\hsize#2\vfill}\vrule}\hrule}\hfill}%
    \else
    {\hfill\epsfbox{#3}\hfill}%
    \fi}

\def\Dp  {\ensuremath{D^+}\xspace}
\def\Dm  {\ensuremath{D^-}\xspace}

\def\pom {\ensuremath{\pm}\xspace}

\begin{document}

 
 \par\vskip 0.8cm

\preprint{\babar-PUB-\BABARPubYear/\BABARPubNumber} 
\preprint{SLAC-PUB-\SLACPubNumber} 

\begin{flushleft}
\begin{tabular}{l}
\babar-PUB-\BABARPubYear/\BABARPubNumber \\
SLAC-PUB-\SLACPubNumber \\
\end{tabular}
\end{flushleft}

\title{
{\large \bf 
Search for \boldmath  $B \rightarrow {J\mskip -3mu/\mskip
-2mu\psi\mskip 2mu} D$ \unboldmath Decays} 
}

%
%
\author{B.~Aubert}
\author{R.~Barate}
\author{D.~Boutigny}
\author{F.~Couderc}
\author{Y.~Karyotakis}
\author{J.~P.~Lees}
\author{V.~Poireau}
\author{V.~Tisserand}
\author{A.~Zghiche}
\affiliation{Laboratoire de Physique des Particules, F-74941 Annecy-le-Vieux, France }
\author{E.~Grauges}
\affiliation{IFAE, Universitat Autonoma de Barcelona, E-08193 Bellaterra, Barcelona, Spain }
% ...
\collaboration{The \babar\ Collaboration}
\noaffiliation

\date{\today}

\begin{abstract}
We report a search for $B \rightarrow {J\mskip -3mu/\mskip
-2mu\psi\mskip 2mu} D$ decays, based on a sample of $124 \times
10^{6}$ $B\overline{B}$ events collected with the BABAR detector at the PEP-II
storage ring of the Stanford Linear Accelerator Center.

No significant signal is found.  We obtain upper limits on the
branching fractions of $1.3\times 10^{-5}$ for $B^0 \rightarrow
{J\mskip -3mu/\mskip -2mu\psi\mskip 2mu} \overline{D}^0$ and $1.2\times
10^{-4}$ for $B^+ \rightarrow {J\mskip -3mu/\mskip -2mu\psi\mskip 2mu}
D^+$ at 90\% confidence level.
\end{abstract}





\maketitle

\begin{figure}
\centering
\includegraphics[width=3in]{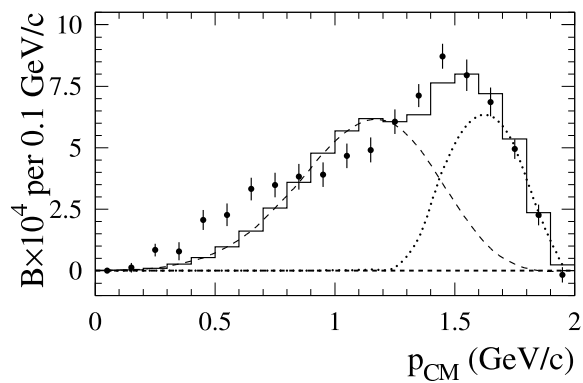}
\caption{Upsilon(4S) rest-frame momentum of \jpsi\ mesons produced  in $B$
decays, after subtracting feed-down 
from $\chi_{c1,2} \to \jpsi \g$ and $\psitwos \to \jpsi \pi\pi$
 (points) \cite{babslowjpsi}.
The histogram is the sum of the color-octet component from a NRQCD
calculation \cite{Beneke:1999gq} (dashed line), which includes
multi-body final states, and the color-singlet $\jpsi K^{(*)}$
component (dotted line).
The
normalization of the curves has been constrained to fit the data.}
\label{fig:directjpsi}
\end{figure}

Measurements of the inclusive spectrum of charmonium mesons in B
decays are in conflict with conventional expectations.
The spectra of the momentum of the \jpsi mesons in the \FourS rest
frame observed by CLEO \cite{ref:cleo} and by BaBar \cite{babslowjpsi}
(Fig.~\ref{fig:directjpsi}), compared with calculations using
non-relativistic QCD (NRQCD) \cite{Beneke:1999gq}, show an excess at
low momentum, corresponding to a branching fraction of approximately
$6 \times 10^{-4}$.
Various hypotheses have been proposed to explain this low-momentum excess.

Brodsky and Navarra \cite{Brodsky:1997yr} have suggested that the
decay $\B\rightarrow \jpsi \Lambda \bar{p}$ \cite{charge}, with the possible
formation of a $\Lambda$-$\bar{p}$ bound state, could explain the CLEO
result.
The kinematic boundary of this structure corresponds to the case where
the \jpsi recoils nearly monoenergetically in the $B$ rest frame
against a $2 \gevcc$ particle.
The $\Lambda$-$\bar{p}$ state could be observed near or just below
threshold.  
\babar\ has searched for these decays and obtained an upper limit of $2.6
\times 10^{-5}$ at 90\% confidence level (C.L.)~\cite{lambdap},
too small to support
the mechanism proposed in \cite{Brodsky:1997yr}.

Decays to a \jpsi meson and a hybrid meson, i.e. a bound state of two
quarks and a gluon, have been proposed~\cite{Davies,ref:Chua}.
In this case the hybrid meson, possibly a $(s\bar{d}g)$ state, would need to
have a mass of about 2 \gevcc.
No experimental evidence has been found to support this mechanism.

If \B mesons were decaying to a narrow resonance and a \jpsi meson,
the \jpsi meson would be monoenergetic in the $B$ rest frame.
Such peaks would appear smeared with an RMS of 0.12\gevc\ in
 Fig.~\ref{fig:directjpsi}, due to the motion of the $B$ in the \FourS rest frame.

The presence of $b\bar{u}c\bar{c}$ components (intrinsic charm) in the
 $B$-meson wave function has also been proposed. In that case the
 charmonium meson is obtained merely by  dissociation when the $b$
 quark decays.
Intrinsic charm was first introduced by Brodsky \textit{et al.}
 \cite{Brodsky:1980pb} to explain an unexpectedly large
cross-section for charmed-particle production in hadron collisions.
Using the estimated amount of intrinsic charm in the proton as an input, 
Chang and Hou predict $\B\to\jpsi D(\pi)$ decays with branching fractions
of the order of  $10^{-4}$~\cite{Chang:2001iy}.
The dominant final state is expected to be $\B\to\jpsi D\pi$, for
which \babar\ 
 has reported an upper limit at 90\% C.L.  of $5.2 \times
10^{-5}$ for $\Bp\to\jpsi\Dz\pip$~\cite{Alessia}.
Four-body decays such as $\B\to \jpsi D\pi\pi$ should be extremely
suppressed by the small phase space available near the kinematical
limit.
The remaining untested final state is $\jpsi D$.
Calculations by Eilam {\em et al.} using perturbative QCD
\cite{Eilam:2001kw} predict branching fractions
(BF's) for  $\B\to \jpsi D$ decays on the order of 
$10^{-8}$ -- $10^{-7}$.
The observation of a signal with a BF significantly larger would
 suggest the presence of intrinsic charm inside the \B meson.

In this Letter we report a search
 for decays  $\B\to\jpsi D$, with  
\Dzb decaying to $\Kp\pim$, \Dp to $\KS\pip$, \KS to $\pip\pim$, and 
  $\jpsi\to\ellell$, where $\ell$ is $e$ or $\mu$.

The data used in this analysis were collected with the \babar\
detector at the \pep2\ storage ring and comprise an integrated
luminosity of 112~\invfb  taken at the \FourS resonance.
The \babar\ detector is described in detail
elsewhere~\cite{ref:babarnim}.  A five-layer, double-sided silicon
vertex tracker (SVT) surrounds the interaction point and provides
precise reconstruction of track angles and \B-decay vertices.
A 40-layer drift chamber (DCH) provides measurements of the transverse
momenta of charged particles.
An internally reflecting ring-imaging Cherenkov detector (DIRC) is
used for particle identification.
A CsI(Tl) crystal electromagnetic calorimeter (EMC)
 detects photons and electrons.
The calorimeter is surrounded by a solenoidal magnet providing a 1.5-T
field. 
The flux return is instrumented with resistive plate chambers 
used for muon and neutral-hadron identification.

We select multihadron events by demanding a minimum of three reconstructed charged tracks in the
polar angle range $0.41 < \theta_{lab} < 2.54$\rad.
A charged track must be reconstructed in the DCH, and,
except for the reconstruction of   $\KS \to \pip\pim$,
it must originate at the nominal interaction point to within 1.5\cm in
the plane transverse to the beam and to within 10\cm along the beam.
Events are required to have an \FourS production point
within 0.5\cm of the
average position of the interaction point in the plane transverse to
the beamline, and  within 6\cm longitudinally.
Neutral clusters  are defined as 
electromagnetic depositions in the calorimeter in the polar angle
range $0.410 < \theta_{lab} < 2.409~\rad$ that are  not associated with
charged tracks and that have an energy greater than 30\mev and 
a shower shape consistent with a photon interaction.
We require the total energy for charged tracks and photon candidates
in the fiducial region to be greater than 4.5\gev.
To reduce continuum $\epem \to \qqbar$
background, we require the ratio of second-to-zeroth Fox-Wolfram moments
 $R_2$~\cite{FoxWolfram} of the event, calculated with both charged
tracks and neutral clusters, to be less than 0.5.
Charged tracks are required to be in regions of polar angle for which
the particle identification (PID) efficiency is well measured.
For electrons, muons, and kaons the acceptable ranges are 0.40 to
2.40, 0.30 to 2.70, and 0.45 to 2.50 rad, respectively.

We further select signal events as described in the following.
Event selection is optimized by maximizing the sensitivity
$s\equiv \epsilon/(a/2+\sqrt{N_B})$, where $a=3$ 
is the number of standard deviations of significance
desired~\cite{punzi}.
The maximum of this ratio is independent of the unknown signal branching fraction.
The signal efficiency $\epsilon$ after all selection requirements is
estimated from simulated Monte Carlo (MC) samples.
The number of background events $N_B$, scaled to the integrated luminosity of the data, is estimated using inclusive
$\FourS\to\BB$ and $\epem\to\qqbar$ MC samples.

We reconstruct \jpsi candidates from a pair of oppositely charged
lepton candidates that form a good vertex.
Muon (electron) candidates are identified with a neural-network
(cut-based) selector. 
For $\jpsi\to\epem$ decays, electron candidates are provisionally combined
with nearby photon candidates in order to recover some of the energy
lost through bremsstrahlung.
These bremsstrahlung-photon candidates are characterized 
by a deposit of more that 30 MeV in the electromagnetic calorimeter
 and a polar angle within 35 mrad of the
 electron direction, as well as an azimuthal angle either within 50
 mrad of the electron direction, or between the electron direction at
 the origin and the azimuth of the impact point in the EMC.
The lepton-pair invariant mass must be in the range [3.00, 3.14] \gevcc for
both lepton flavors. 

We form \KS candidates from oppositely charged tracks originating from
a common vertex and having an invariant mass in the range [487, 510] \mevcc.
The \KS flight length must be greater than 1\mm, and its direction in
the plane perpendicular to the beam line must be within 0.2\rad of the
\KS momentum vector.
All charged tracks are taken as pion candidates, and kaon candidates
are identified with a likelihood selector based on Cherenkov-angle
measurements from the DIRC and specific ionization in the SVT and in
the DCH.
Candidates for $D$ mesons
 are formed from $K\pi$ combinations; a requirement on the 
 $K\pi$ invariant mass $m_{K\pi}$ is applied during the optimization of the selection. 
The analysis is then performed in a larger window   $1.80 < m_{K\pi} < 1.92
\gevcc$.
The high statistics decays \B\to \jpsi \Kstar  with the same
$\jpsi K\pi$ final state are used as a control sample to evaluate the
possible differences between  data and  MC.
These are selected with requirements similar to those of the signal,
except for an $m_{K\pi}$ range of [0.79, 0.99] \gevcc.
The \jpsi and \KS candidates are constrained to their
 nominal masses~\cite{ref:pdg} to improve the resolution
of the measurement of the four-momentum of their parent-\B candidate.

Candidate \B mesons are formed from \jpsi and $D$ candidates.
Two kinematic variables are used to further remove incorrectly reconstructed \B candidates.
The first is the difference $\Delta E \equiv E^*_B - E^*_{beam}$ between
the \B-candidate energy and the beam energy in the \FourS rest
frame. In the absence of experimental effects, correctly reconstructed signal
candidates have $\Delta E = 0$. 
The  $\Delta E$ resolution is  7.5 \mev.
For the signal region, $\Delta E$ is required to be in the range
$[-15, +12] \mev$.
The second variable is the energy-substituted mass $\mes \equiv
(E^{*2}_{beam} -p^{*2}_B)^{1/2}$, where $p^{*}_B$ is the momentum of the
\B candidate in the \FourS rest frame.
The energy substituted mass \mes  peaks at the nominal \B  mass of 5.279\gevcc for the signal.
Its typical resolution is 2.5 \mevcc.
A requirement of $5.274<\mes< 5.284 \gevcc$ was obtained in the
optimization of the signal selection.
The analysis is then performed in the window $5.2<\mes<5.3 \gevcc$.
If more than one \B candidate is found in an event, the one having the
smallest $|\DeltaE|$ is retained.

\begin{figure}[h]
\begin{center}
\includegraphics[width=0.95\linewidth]{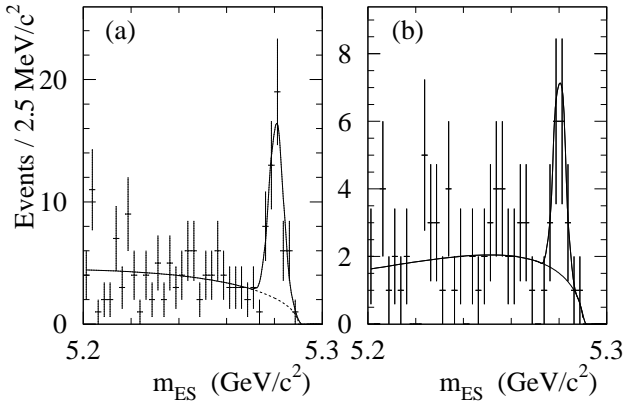}
\end{center}
\caption{Distribution of  \mes  
for data events with   $m_{K\pi}$ in the range  [1.8, 1.92] \gevcc.
(a): $\jpsi\Dzb$, (b): $\jpsi\Dp$.
The fits are described in the text.
The peaks at the nominal \B mass are due to non-$D$ events.
\label{fig:Argus:data}}
\end{figure}

Non-$D$  $\B\to\jpsi K\pi$ decays that have \mes, $\Delta E$,
and $m_{\ellell}$ distributions
similar to those of the signal are found
 to be the dominant contribution to the remaining
background after selection cuts are applied.
Signal events can be separated from non-$D$ events by their peaking at the
$D$ invariant mass in the $m_{K\pi}$ spectrum.
In MC samples, this spectrum shows a small
 but significant number of true $D$ mesons: a  $D$ meson from the
 decay of one $B$ was combined with a  \jpsi meson from the decay
 of the other \B.
We subtract this combinatorial background using the \mes distribution:
 the $m_{K\pi}$ distribution of the events in the sideband
($5.21<\mes<5.27 \gevcc$) is subtracted from the distribution of the
events in the signal region, 
with a scaling factor $R$ that is the ratio of the combinatorial
background in the signal region and in the sideband.
The value of $R$ is obtained from  the integrals of the ARGUS
 shape~\cite{argus} in fits of the \mes distribution with a Gaussian
 function for the signal and an ARGUS shape for the combinatorial
 background (Fig.~\ref{fig:Argus:data}).
The \mes signal window for the data is shifted by +0.6 \mevcc with respect
to MC after such a shift is observed on the \jpsi\Kstar control
sample.
We obtain $R = 0.093 \pom 0.011$ and $R = 0.131 \pom 0.038$ for
$\jpsi\Dzb$ and $\jpsi\Dp$, respectively.

\begin{figure}[h]
\begin{center}
\includegraphics[width=0.95\linewidth]{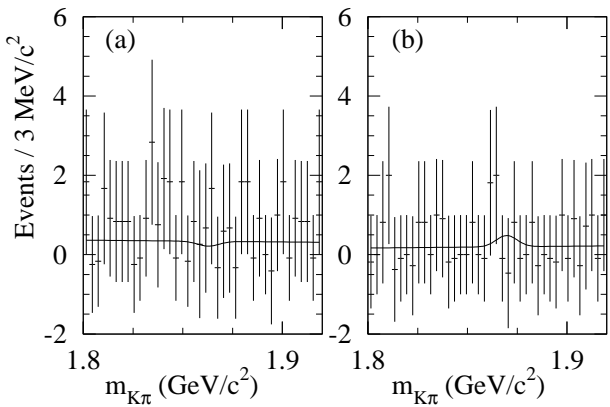}
\end{center}
\caption{Background-subtracted
 $m_{K\pi}$ distributions.
((a): $\jpsi\Dzb$, (b): $\jpsi\Dp$)
The fits are described in the text.
\label{fig:mkpi:data}}
\end{figure}

The combinatorial-background-subtracted $m_{K\pi}$ distributions
(Fig.~\ref{fig:mkpi:data}) are fitted with the combination of a linear
background (two free parameters) and a Gaussian signal in which the
number of signal events $S$ is a free parameter.
The central value and the resolution of the $D$ peak are fixed in the
fit.
The resolution measured on the signal MC sample is used.
The agreement of data and MC samples has been studied with a $D$-meson
 control sample obtained with the same selection as for the $D$
 candidates of the $\jpsi D$ events.
The resolutions are found to be similar in the data and in the
MC samples, and a shift of $-0.6 \pom 0.2 \mevcc$ of the central value
 is observed and accounted for.

No significant signal for $\B\to\jpsi D$ is observed.  
The numbers of events obtained are $-0.6 \pom 1.2 \pom 0.2$ ($\jpsi\Dzb$) and
$1.2 \pom 1.9 \pom 0.2$ ($\jpsi\Dp$), where the first uncertainty is
statistical, and the second one is the systematic contribution due to the
uncertainties in the scaling factor for background subtraction, and
of the $D$ mass and mass resolution used in the fit.
The branching fractions are:
\begin{equation}
\label{eq:BF:data}
\BR = \frac{S}{N_{evt} \times \epsilon \times b },
\end{equation}

\noindent
where $S$ is the number of signal events obtained from the fit,
 $N_{evt}=124\times 10^{6}$ is the number  of \BB events in the data sample, 
and $b$ is the product of the branching fractions of the secondary 
decays (Table \ref{tab:res:data}).

\begin{table}[h]
\caption{Number of signal events, efficiency, secondary branching fraction, 
measured branching fraction (\BR) and upper limit (UL) at  90\% C.L.
\label{tab:res:data}
}
\begin{center}
\begin{tabular}{lccccccc}
 \hline
 \hline
& $S$ & $\epsilon$ & $b$ &\BR & UL  \\
& & (\%) & ($10^{-3}$) & ($10^{-5}$) &  ($10^{-5}$)\\
 \hline
 \jpsi \Dzb & $-0.6$ \pom 1.2 & 23.3 \pom 0.3 &  4.49 &  $-0.46$ \pom  0.93  &   1.3 \\
 \jpsi \Dp & 1.2 \pom 1.9 & 22.6 \pom 0.2 & 1.15 &   3.7 \pom  5.9  &  12.3 \\
 \hline
 \hline
\end{tabular}
\end{center}
\end{table}

Additional contributions to the systematic uncertainty of the
branching fraction are described in the following.
The relative uncertainty in the number of \BB events is 1.1\%.
The secondary branching fractions and their uncertainties are taken from
 PDG~\cite{ref:pdg}.
Other estimated uncertainties are from tracking efficiency (1.3\% per
 track added linearly), \KS reconstruction (2.5\%), PID efficiency
 (3.0\%) and 
the statistical uncertainty in the selection efficiency. 
The uncertainty in the selection efficiency due to the uncertainty of
the MC/data difference of the central value and of the width of the
peaks in $m_{\ellell}$, \mes, and $\Delta E$ is estimated from the
\jpsi \Kstar control sample.
A summary of the multiplicative contributions to the systematics can
be found in Table\,\ref{tab:SumSyst}.
The ratio of \Bz to \Bp production in \FourS decays is assumed to be unity.
The related uncertainty is small and is neglected here.

\begin{table}[htbp]
\caption{Summary of the contributions to the relative systematic
uncertainty
(\%).
\label{tab:SumSyst}}
\begin{center} 
\begin{tabular}{lcc}
\hline 
\hline 
& $\jpsi\Dzb$ (\Kp \pim) & $\jpsi\Dp$ (\KS \pip) \\
\hline 
$B$ counting & 1.1 & 1.1 \\
Secondary BF's & 2.7 & 6.8 \\
Tracking & 5.2 & 3.9 \\
\KS & -- & 2.5 \\
PID & 3.0 & 3.0 \\
MC statistics & 1.5 & 1.0 \\
Sample selection & 1.0 & 0.8 \\
\hline 
Total & 6.9 & 8.9 \\
 \hline 
\hline 
\end{tabular}
\end{center}
\end{table}

We obtain upper bounds on the branching fractions at 90\% confidence
level (C.L.)  assuming Gaussian statistics for the statistical
uncertainties and taking into account the systematic uncertainties.
We have used a Bayesian method with uniform prior for positive BF
values in the derivation of these limits.
We obtain upper limits 
of $1.3\times 10^{-5}$ for \Bz \to \jpsi \Dzb and
$1.2\times 10^{-4}$ for \Bu \to \jpsi \Dp.
The results for the neutral decay are significantly lower than the
amount needed to explain the excess of low momentum \jpsi's after
Chang and Hou \cite{Chang:2001iy}.
Together with the small upper limits on the branching fraction for decays
$\B \to \jpsi D \pi$~\cite{Alessia},
 we  conclude that intrinsic charm as the explanation
of the low momentum \jpsi excess in \B decays  is not supported.

We are grateful for the excellent luminosity and machine conditions
provided by our \pep2\ colleagues, 
and for the substantial dedicated effort from
the computing organizations that support \babar.
The collaborating institutions wish to thank 
SLAC for its support and kind hospitality. 
This work is supported by
DOE
and NSF (USA),
NSERC (Canada),
IHEP (China),
CEA and
CNRS-IN2P3
(France),
BMBF and DFG
(Germany),
INFN (Italy),
FOM (The Netherlands),
NFR (Norway),
MIST (Russia), and
PPARC (United Kingdom). 
Individuals have received support from CONACyT (Mexico), A.~P.~Sloan Foundation, 
Research Corporation,
and Alexander von Humboldt Foundation.



\begin{thebibliography}{99}
\bibitem{charge}
Charge-conjugate modes are included implicitly throughout this paper.
\end{thebibliography}
\end{document}